\documentclass[a4paper,fleqn]{cas-dc}
\usepackage[numbers]{natbib}
\usepackage{amsfonts}
\usepackage{algorithmic}
\usepackage{amsmath} 
\usepackage{amssymb} 
\usepackage{caption}
\usepackage{subcaption}
\usepackage{etoolbox}
\usepackage{forest}
\usepackage{lingmacros}
\usepackage{textcomp}
\usepackage{tree-dvips}
\usepackage{tikz}
\usepackage{tikz-cd}
\usepackage[arrowdel]{physics}
\usepackage{graphicx}
\usepackage{wrapfig}
\usepackage{listings}
\usepackage{pgfplots, pgfplotstable}
\usepackage{diagbox} 
\usepackage[usestackEOL]{stackengine}
\usepackage{makecell}
\usepackage{mathrsfs}
\usepackage{moresize}
\usepackage{multirow}
\usepackage{multicol}
\usepackage[numbers]{natbib}
\usepackage[T1]{fontenc}
\usepackage{xcolor}
\allowdisplaybreaks[1]
\definecolor{orchid}{rgb}{0.7, 0.4, 1.1}
\definecolor{comment_color}{rgb}{0, 0.5, 0}
\definecolor{keyword_color}{rgb}{0.3, 0, 0.6}
\definecolor{string_color}{rgb}{0.5, 0, 0.1}

\begin{document}
\shorttitle{Organic liquid scintillator neutrino detector experiment, theoretical modeling, and computational simulation}

\shortauthors{Xi Liu$^a$, Wenxi Fang$^b$}
\title{{\Large Organic liquid scintillator neutrino detector experiment, theoretical modeling, and computational simulation}}
\author[1]{\color{black}Xi Liu}
\author[2]{\color{black}Wenxi Fang}

\address{$^a$xl3467@columbia.edu, Columbia University; $^b$u3013972@connect.hku.hk, Inner Mongolia University of Science and Technology}
\begin{abstract}
The Liquid Scintillator Neutrino Detector (LSND) experiment aimed at investigating neutrino oscillations, particularly the transformation of muon-type antineutrinos (\(\overline{\nu}_\mu\)) into electron-type antineutrinos (\(\overline{\nu}_e\)). This phenomenon challenges the Standard Model's assumption of massless neutrinos. The LSND employed a large organic liquid scintillator (LS) to detect low-energy neutrino interactions, enhanced by the addition of metal ions such as gadolinium (Gd) for improved signal sensitivity and noise suppression. Theoretical modeling and simulations are used in this paper to accurately interpret experimental results. The FLUKA Monte Carlo code was employed to simulate particle interactions and transport in the detector. Key processes modeled included neutrino interactions (\(\overline{\nu}_e + p \to e^+ + n\)), neutron capture (\(n + p \to d + \gamma\)), and the corresponding light output in the scintillator. The simulations accounted for quenching effects using Birks' law, enabling precise predictions of light yield and detector response to secondary particles. Neutrino fluxes from decay-at-rest (DAR) and decay-in-flight (DIF) processes were calculated, capturing the energy spectra of neutrinos generated by pion and muon decays. Pion production cross-sections and light output efficiency for various particles were also modeled to understand detector performance comprehensively. The theoretical modeling and simulation framework validated the experimental observations and provided insights into the detector's sensitivity and limitations. The LSND results hinted at deviations from the Standard Model, stimulating further investigations into neutrino oscillations and the potential existence of sterile neutrinos.
\end{abstract}


\maketitle
\section{Introduction}
The liquid scintillator neutrino detector (LSND) experiment was designed to investigate neutrino oscillations by detecting neutrinos produced at the Los Alamos meson physics facility (LAMPF). The experiment searched for the conversion of muon-type antineutrinos (\(\overline{\nu}_{\mu}\)) into electron-type antineutrinos (\(\overline{\nu}_e\)), i.e., $\overline{\nu}_\mu\rightarrow\overline{\nu}_e$.

Neutrinos are fundamental particles in the Standard Model, belonging to the lepton family. Neutrinos interacts only via the weak nuclear force, so they are difficult to detect. Neutrinos have three flavors: electron neutrinos (\(\nu_e\)), muon neutrinos (\(\nu_\mu\)), and tau neutrinos (\(\nu_\tau\)). The conservation of lepton number is an important principle in particle physics, ensuring that the total number of leptons minus antileptons remains the same in a particle reaction.

Neutrino oscillation is a phenomenon in which neutrinos change from one flavor to another as they travel through space. This process occurs because neutrino flavor states (which we observe) are quantum superpositions of mass states (which propagate in space). If neutrinos have mass, they can oscillate from one flavor to another. This is a departure from the Standard Model, which initially predicted massless neutrinos. LSND specifically looked for the oscillation of muon neutrinos into electron neutrinos, a signature that would indicate the existence of neutrino oscillations and suggest new physics beyond the Standard Model.

Organic liquid scintillators have been widely used in past and current neutrino physics experiments, particularly for detecting low-energy neutrinos where real-time data and precise energy measurements are needed \cite{athanassopoulos_1997}. These scintillators offer several advantages over other detection technologies. To enhance the neutrino signal and reduce background noise, metals are often added to the organic liquid. However, in the past, many metal-loaded scintillators faced issues with chemical and optical instability, which impacted the performance of neutrino detectors.
There are various methods of metal loading, with an emphasis on recent techniques that produce stable metal-loaded scintillators, suitable for long-term use in large-scale experiments, and also covers the applications of metal-loaded scintillators in neutrino research, comparing the performance and future potential of different scintillator types.

The addition of metal ions to liquid scintillators (LSs) offers significant benefits for detecting weak interaction signals, such as positrons and neutrons generated by inverse beta decay (IBD) \cite{buck_2016}. Metals with high thermal-neutron cross sections, including $^{108}$Cd, $^6$Li, and $^{157}$Gd, enhance delayed signal discrimination compared to hydrogen, which has a much lower cross section of approximately 0.3 b. Despite these advantages, early metal-loaded LSs encountered issues such as poor stability and short attenuation lengths, as previously noted.

Liquid scintillators (LSs) must be large-scale and maintain long-term stability for effective neutino detection, as neutrino interactions with matter are exceedingly rare. Essential characteristics of LSs include high optical transparency, radiopurity, and durable optical and chemical stability. Incorporating metals into LSs presents significant challenges, primarily due to the difficulty of integrating polar metal ions into nonpolar LS solvents without degrading scintillator performance. The disparity in densities and polarities between metal ions and LS solvents complicates the achievement of uniform dispersion.

One commonly used method is the synthesis of organometallic compounds by coordinating metals with ligands. Ligands play a crucial role in resolving dispersion issues by enabling the uniform incorporation of metal ions into the LS matrix \cite{choi_2024}. Traditional ionic compounds typically have poor solubility in hydrophobic organic solvents due to their hydrophilic characteristics. Organic ligands convert these metal ions into neutral, soluble complexes that are compatible with solvent extraction, achieved through heterogeneous chemical reactions at the aqueous–organic interface.

\begin{figure}
\centering
\caption{solvent candidates density, flashpoint, and wavelengths of the optical absorption/emission peaks (when dissolved in cyclohexane)}
\includegraphics[width = 0.5\textwidth, height = 0.2\textwidth]{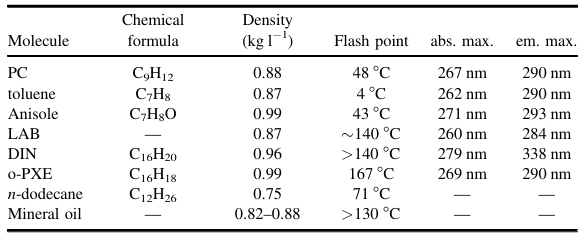}
\end{figure}
\begin{figure}
\centering
\caption{primary and secondary fluor molecule wavelengths of the optical absorption and emission peak maxima (when diluted in cyclohexane)}
\includegraphics[width = 0.45\textwidth, height = 0.2\textwidth]{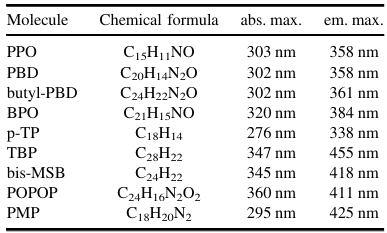}
\end{figure}

Ligands containing functional groups such as carboxylic acids, beta-diketones, and crown ethers can donate oxygen atoms to form complexes with metal ions \cite{back_2008}. Carboxylic acid complexes are widely employed in contemporary neutrino-detection experiments. Beta-diketones are especially effective due to their strong hydrophobic properties, which make them highly efficient for extracting metal ions in organic solvents. In contrast, chelating agents such as crown ethers typically exhibit lower efficiency in metal ion extraction compared to organic ligands.

The metal-loaded LSs has been used through numerous key neutrino experiments. Metals such as Gd, Cd, and Li have been tested to optimize signal signatures while suppressing background noise \cite{boubker_1989}. A notable success in this field is the Reactor Experiment for Neutrino Oscillation (RENO) in South Korea. Despite initial challenges with Gd-loaded LSs, RENO demonstrated that Gd-loaded liquid scintillators (GdLSs) can maintain stability for over five years after synthesis. The synthesis of GdLSs typically involves forming Gd organometallic complexes using carboxylic acids through a neutralization reaction, as shown:
\begin{align*}
&RCOOH + NH_3 \rightarrow RCOONH_4 + H_2O\\
&3RCOONH_4 + GdCl_3 \rightarrow Gd(RCOO)_3 + 3NH_4Cl
\end{align*}
The carboxylic acid precursor is 3,5,5-trimethylhexanoic acid (TMHA), neutralized with ammonium hydroxide. One of the predominant methods for incorporating metal carboxylates into scintillators in neutrino experiments is the liquid-liquid extraction technique.

The metal complex precursor GdCl$_3$ is combined with ligands, such as 3RCOONH$_4$, in two immiscible solvents, typically an organic phase and an aqueous phase in liquid-liquid extraction. The Gd and ligand solutions are introduced as small droplets, allowing them to mix within the solvents and transition from a turbulent to a laminar flow region. The immiscible solvents then separate into a nonpolar organic layer and a polar aqueous layer, with the precursor partitioning between the two. Ion-exchange processes involving adsorption and diffusion facilitate the interchange of ions in the solution.
\begin{figure}
\centering
\caption{Inverse beta decay in the detector}
\includegraphics[width = 0.25\textwidth, height = 0.25\textwidth]{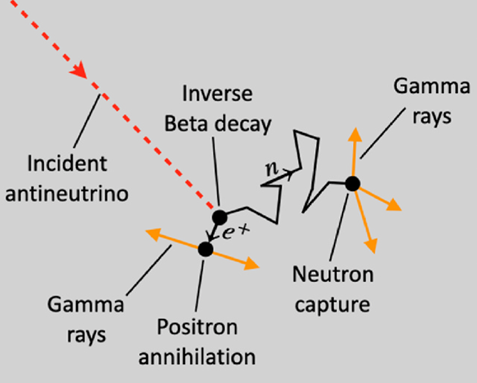}
\end{figure}
\begin{figure}
\centering
\caption{Daya bay neutrino detector}
\includegraphics[width = 0.25\textwidth, height = 0.25\textwidth]{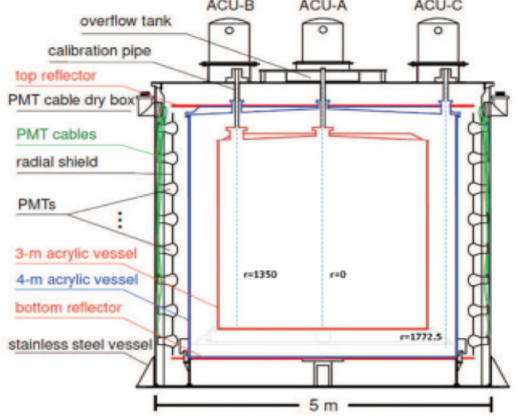}
\end{figure}

In the reactor experiment for neutrino oscillation (RENO), the liquid–liquid extraction technique has produced pure, transparent, and stable metal-loaded LSs. This extraction process enables the transfer of Gd from an aqueous phase to an organic phase. A neutralization solution of TMHA and an aqueous solution of GdCl$_3$ are combined with a LAB solution. Through neutralization, a Gd complex (GdR$_3$) is formed in the aqueous phase, which subsequently dissolves in the LAB solution due to its greater solubility in the organic solvent than in water. Following the reaction, the density difference between the aqueous and LAB phases causes the solutions to separate. The lower aqueous layer is removed, leaving behind the Gd-loaded LAB, which is then subjected to N$_2$ purging to enhance the attenuation length. This method produces a stable GdLS with excellent attenuation lengths. Similar procedures can be used to incorporate other metals, such as boron (B) and lithium (Li).

A significant development in organic LSs is the incorporation of di-isopropyl naphthalene (DIN). DIN offers the advantage of improving the discrimination between signal and background events in neutrino experiments, particularly those involving fast neutrons and gamma rays. The addition of DIN modifies the pulse shapes of "neutron-like" and "gamma-like" events, enhancing the performance of pulse-shape discriminators. This improvement helps distinguish true neutrino interactions from background events, such as those caused by radioactive decay or environmental noise. The enhanced discrimination provided by DIN-loaded scintillators is especially valuable for reducing false signals, a persistent challenge in neutrino physics. Preliminary studies indicate a significant improvement in the signal-to-noise ratio, making DIN-loaded LSs highly suitable for experiments requiring effective background suppression, particularly in short-baseline neutrino studies.
\section{Neutrino oscillation}
The flavor eigenstates form a complete orthonormal basis. Another basis, composed of mass eigenstates $\nu_1$, $\nu_2$, and $\nu_3$, diagonalizes the free-particle Hamiltonian. Neutrino oscillation experiments confirm that these two bases are rotated relative to each other. Each flavor eigenstate is a superposition of mass eigenstates, described by the PMNS matrix $U_{\alpha i}$, which parameterizes the unitary transformation between the bases:

\begin{align*}
\begin{pmatrix}
\nu_e \\
\nu_\mu \\
\nu_\tau
\end{pmatrix}
=
\begin{pmatrix}
U_{e1} & U_{e2} & U_{e3} \\
U_{\mu1} & U_{\mu2} & U_{\mu3} \\
U_{\tau1} & U_{\tau2} & U_{\tau3}
\end{pmatrix}
\begin{pmatrix}
\nu_1 \\
\nu_2 \\
\nu_3
\end{pmatrix}
\end{align*}

A neutrino in a flavor state $\alpha$ is a mixture of mass states. The probability of observing mass $m_i$ is $|U_{\alpha i}|^2$. Under CPT symmetry, the PMNS matrix for antineutrinos is identical to that for neutrinos. Compared to quarks, determining PMNS coefficients is more challenging due to the difficulty of detecting neutrinos.

The PMNS matrix is a unitary $3 \times 3$ matrix parameterized by three mixing angles ($\theta_{12}$, $\theta_{23}$, $\theta_{13}$) and one CP-violating phase $\delta_{\mathrm{CP}}$. This parameterization is
{\small
\begin{align*}
U =
\begin{pmatrix}
1 & 0 & 0 \\
0 & c_{23} & s_{23} \\
0 & -s_{23} & c_{23}
\end{pmatrix}
\begin{pmatrix}
c_{13} & 0 & s_{13}e^{-i\delta_{\mathrm{CP}}} \\
0 & 1 & 0 \\
-s_{13}e^{i\delta_{\mathrm{CP}}} & 0 & c_{13}
\end{pmatrix}
\begin{pmatrix}
c_{12} & s_{12} & 0 \\
-s_{12} & c_{12} & 0 \\
0 & 0 & 1
\end{pmatrix}
\end{align*}}
where $s_{ij} = \sin\theta_{ij}$ and $c_{ij} = \cos\theta_{ij}$. For Majorana neutrinos, two additional complex phases, $\alpha$ and $\beta$, are required, as the phases of Majorana fields are constrained by $\nu = \nu^c$. The diagonalized mass matrix $M_\nu$ is related to the mixing matrix by $M_\nu = U M_D U^\dagger$, where $M_D$ is the diagonalized mass matrix.

The mixing angles have been measured: $s^2_{12} = 0.31$, $s^2_{23} = 0.42$, $s^2_{13} = 0.024$, with mass-squared differences $\Delta m^2_{21} = 7.5 \times 10^{-5} \, \text{eV}^2$ and $|\Delta m^2_{31}| = 2.47 \times 10^{-3} \, \text{eV}^2$. The CP-violating phase $\delta_{\mathrm{CP}}$ remains unmeasured, though fits provide estimates \cite{fukumi_2012}. Open questions include the nature of neutrino masses (Dirac or Majorana), the smallest mass $m_0$, and CP-violating phases $\alpha$, $\beta$, and $\delta$. Experimental challenges arise in measuring the effective neutrino mass, particularly through neutrino mass spectroscopy using atomic or molecular targets.

Consider the particle interaction \( a+b \to c+d \), where the exchange of an intermediate particle \( X \) occurs. In perturbation theory, the transition amplitude \( T_{fi} \) for a process is
\[
T_{fi} = \langle f | V | i \rangle + \sum_{j} \frac{\langle f | V | j \rangle \langle j | V | i \rangle}{E_i - E_j} + \cdots
\]
The initial state is \( |i\rangle = |a+b\rangle \) with energy \( E_i = E_a + E_b \). The intermediate state is \( |j\rangle = |c + b + X\rangle \) with energy \( E_j = E_c + E_X + E_b \). The final state is \( |f\rangle = |c + d\rangle \) with energy $E_c+E_d$. The intermediate state includes the exchanged particle \( X \). The second order term is
\[
T_{fi} = \sum_{j} \frac{\langle f | V | j \rangle \langle j | V | i \rangle}{E_i - E_j}
\]
The first transition involves the interaction \( |a + b\rangle \to |c + b + X\rangle \), with matrix element \( \langle j | V | i \rangle = \langle c + b + X | V | a + b \rangle \).\\
The second transition involves \( |c + b + X\rangle \to |c + d\rangle \), with matrix element \( \langle f | V | j \rangle = \langle c + d | V | c + b + X \rangle \)
\begin{align*}
&T_{fi} = \frac{\langle f | V | j \rangle \langle j | V | i \rangle}{E_i - E_j}\\
&=\frac{\langle c + d | V | c + b + X \rangle \langle c + b + X | V | a + b \rangle}{(E_a + E_b) - (E_c + E_X + E_b)}\\
&= \frac{\langle d | V | X + b \rangle \langle c + X | V | a \rangle}{E_a - (E_c + E_X)}
\end{align*}
The expression for lorentz invariant matrix element \( M\) in quantum electrodynamics is a result of applying Feynman rules to the QED interaction Lagrangian.
\[
M = -ie \bar{u}(p_3) \gamma^\mu u(p_1) \frac{g_{\mu\nu}}{q^2} \bar{u}(p_4) \gamma^\nu u(p_2)
\]
In the weak interaction, the matrix element for a process involving fermions is
\[
M_{fi} = \frac{G_F}{\sqrt{2}} \left[ \bar{\psi}_3 \gamma^\mu (1 - \gamma_5) \psi_1 \right] \left[ \bar{\psi}_4 \gamma_\mu (1 - \gamma_5) \psi_2 \right]
\]
\( G_F \) is the Fermi constant, \( \gamma^\mu \) are the Dirac gamma matrices, \( \gamma_5 \) is the chirality matrix that enforces the parity violation in the weak interaction, \( \psi_1, \psi_2 \) are the fermions involved in the initial state, and \( \psi_3, \psi_4 \) are the fermions involved in the final state.

For most low-energy weak interactions, we approximate the W-boson propagator. The propagator for the W boson in the limit \( |q^2| \ll m_W^2 \) (where \( q^2 \) is the four-momentum transfer and \( m_W \) is the mass of the W boson) is
\[
\frac{-i g_{\mu\nu}}{q^2 - m_W^2} \approx \frac{-i g_{\mu\nu}}{m_W^2}
\]
where \( g_{\mu\nu} \) is the metric tensor for the spacetime index.

Fermi originally described the weak interaction as a contact interaction (which ignores the propagator, assuming \( q^2 \) is large), with the matrix element:
\[
M_{fi} = \frac{G_F}{\sqrt{2}} \, g_{\mu\nu} \left[ \bar{\psi}_3 \gamma^\mu \psi_1 \right] \left[ \bar{\psi}_4 \gamma^\nu \psi_2 \right]
\]
This contact interaction represents a local interaction, where the W boson is exchanged instantaneously at the interaction point, thus effectively not involving a propagator.

The weak interaction violates parity, so modify the gamma matrices at each vertex to include the \( (1 - \gamma_5) \) term, which enforces left-handed chirality. The modified matrix element becomes
\[
M_{fi} = \frac{G_F}{\sqrt{2}} \, \left[ \bar{\psi}_3 \gamma^\mu (1 - \gamma_5) \psi_1 \right] \left[ \bar{\psi}_4 \gamma_\mu (1 - \gamma_5) \psi_2 \right]
\]

This modification is due to that only left-handed fermions participate in the weak interaction, reflecting the violation of parity symmetry. The propagator for the W boson is introduced in the intermediate state. The full interaction involves an exchange of a virtual W boson, and the propagator term.
\[
M_{fi} = \frac{G_F}{\sqrt{2}} \, \frac{g_{\mu\nu}}{m_W^2} \left[ \bar{\psi}_3 \gamma^\mu (1 - \gamma_5) \psi_1 \right] \left[ \bar{\psi}_4 \gamma^\nu (1 - \gamma_5) \psi_2 \right]
\]

The weak leptonic current associated with the \( \nu \) vertex in weak interactions is
\[
j^\nu = \frac{g_W}{\sqrt{2}} \, u(p_3) \gamma^\nu (1 - \gamma_5) v(p_4)
\]
Let \( f_\pi p_\mu^\pi \) represent a pion factor, which comes from the strong interaction part (in processes such as pion decay), can write the final matrix element for the decay or scattering process as:
\begin{align*}
&M_{fi} = \frac{g_W}{\sqrt{2}} \, f_\pi p_\mu^\pi \, \frac{g_{\mu\nu}}{m_W^2} \, \frac{g_W}{\sqrt{2}} \, u(p_3) \gamma^\nu (1 - \gamma_5) v(p_4)\\
&= \frac{g_W^2}{4 m_W^2} g_{\mu\nu} f_\pi p_\mu^\pi u(p_3) \gamma^\nu (1 - \gamma_5) v(p_4)
\end{align*}
The matrix element \( M_{fi} \) represents the amplitude for the transition from the initial state to the final state in a particle decay. The decay rate, \( \Gamma \), is related to the squared modulus of the matrix element and the phase space of the decay products. For a process like \( a \rightarrow b + c \), the decay rate is
\[
\Gamma = \frac{1}{2m_a} \int |\mathcal{M}|^2 d\Pi_n
\]

where \( m_a \) is the mass of the decaying particle, \( \mathcal{M} \) is the matrix element for the decay (in this case, \( M_{fi} \)), \( d\Pi_n \) is the differential phase space element for the final state particles.

The phase space element \( d\Pi_n \) is integrated over all allowed final states. For a 2-body decay, the phase space integral takes a simple form, depending on the momenta of the outgoing particles. The phase space factor is
\[
d\Pi_2 = \frac{|\mathbf{p}_b|}{8\pi m_a^2} d\Omega
\]
where \( \mathbf{p}_b \) is the momentum of the final particle in the rest frame of the decaying particle, and \( d\Omega \) is the solid angle element. For more complicated decays (with more than two final particles), the phase space integration becomes more complex and often requires numerical methods or approximations. The decay rate gives a measure of how quickly a particle decays into a particular final state. For weak decays, it is often associated with the interaction strength (in terms of the weak coupling constant \( g_W \)), the mass of the decaying particle, and the phase space available for the decay products.

\section{Particle reactions}
\begin{figure}
\centering
\caption{Kamioka liquid scintillator antineutrino detector}
\includegraphics[width = 0.35\textwidth, height = 0.23\textwidth]{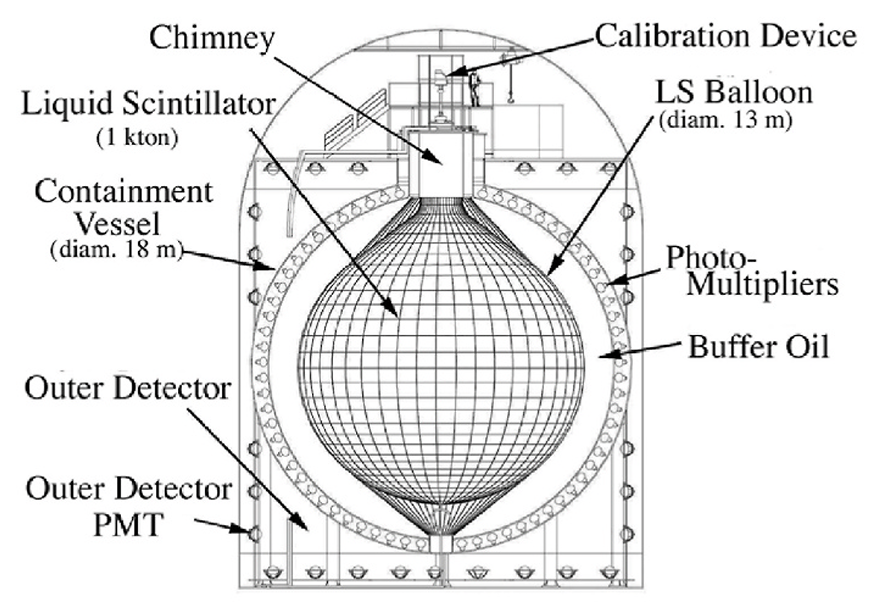}
\end{figure}
\begin{figure}
\centering
\caption{detector and target area}
\includegraphics[width = 0.5\textwidth, height = 0.23\textwidth]{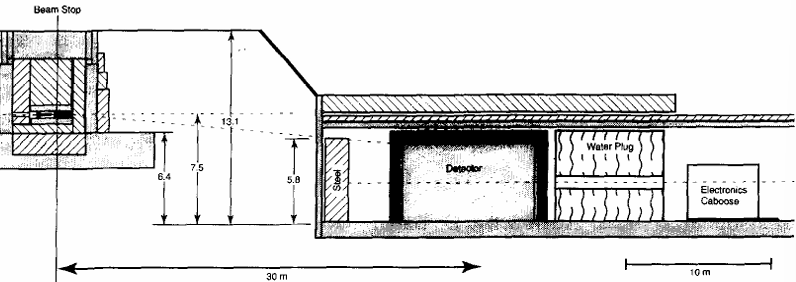}
\end{figure}

At LAMPF, neutrinos were produced by a proton beam hitting a water target, generating pions. Pions (\(\pi^\pm\)) decay into muons (\(\mu^\pm\)) and neutrinos:\\
Positive pion decay: 
\[
\pi^+ \rightarrow \mu^+ + \nu_\mu
\]
Muon decay: 
\[
\mu^+ \rightarrow e^+ + \nu_e + \bar{\nu}_\mu
\]

Most of the positive pions came to rest and decayed at rest, producing muon neutrinos (\(\nu_\mu\)) with a maximum energy of 52.8 MeV \cite{xie_2021}. These neutrinos then traveled to the detector, where the LSND experiment aimed to detect any appearance of electron neutrinos (\(\nu_e\)), which could indicate oscillation.

LSND searched for neutrino oscillations using two complementary techniques: decay at rest (DAR) and decay in flight (DIF). Decay at rest occurs when a positive muon ($\mu^+$) comes to a complete stop within the beam stop before decaying, releasing a muon neutrino ($\nu_\mu$). Decay in flight happens when a pion ($\pi^+$) decays while still moving through the beamline, producing a muon neutrino ($\nu_\mu$) with higher energy due to the pion's motion. Decay at rest produces a lower energy neutrino spectrum compared to decay in flight, as the muon is stationary and all its energy is distributed in the decay products.

\begin{figure}
\centering
\caption{Feynman diagram of inverse beta decay}
\includegraphics[width = 0.3\textwidth, height = 0.24\textwidth]{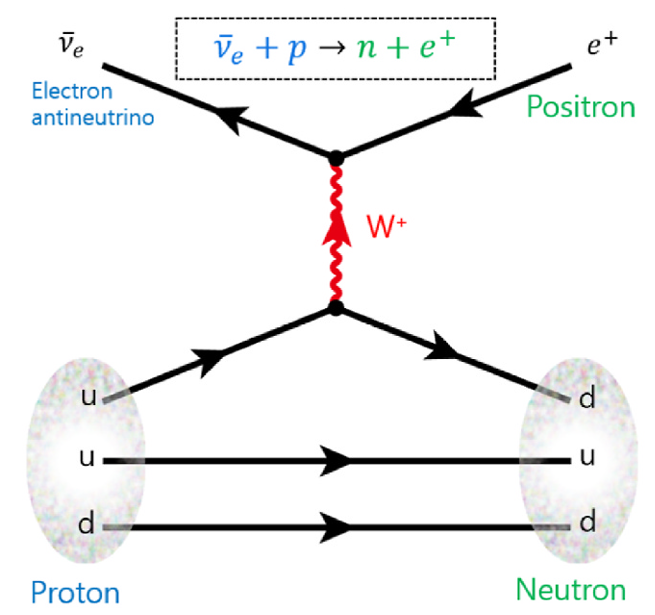}
\end{figure}

In the DAR mode, most of the pions decayed at rest in the target, leading to a well-known flux of muon neutrinos (\(\nu_\mu\)) with a distinct energy spectrum. These neutrinos interacted with the mineral oil in the detector. The signature sought in the DAR mode was the appearance of electron antineutrinos (\(\overline{\nu}_e\)) interacting with protons in the detector via the reaction:
\begin{align*}
&\nu_e + p \rightarrow e^- + n\\
&\nu_e + ^{12}C \rightarrow e^- + ^{12}B\\
&\overline{\nu}_e + p \rightarrow e^+ + n\\
&\overline{\nu}_e + ^{12}C \rightarrow e^+ + ^{12}B
\end{align*}
Carbon and carbon-12 have 6 protons and 6 neutrons. Boron has 5 protons and 6 neutrons, boron-12 has 5 protons and 7 neutrons. The carbon had lost a proton and gained a neutron.

In the neutron-capture reaction:  

\[
n + p \rightarrow d + \gamma
\]

The \(d\) represents deuterium, also known as heavy hydrogen. It is an isotope of hydrogen with one proton and one neutron in its nucleus. A neutron (\(n\)) is captured by a proton (\(p\)) in the medium, such as a liquid scintillator or water, resulting in the formation of deuterium (\(d\)). This process releases a 2.2 MeV gamma photon, which is a distinct detection signature for this reaction.

This reaction is significant in neutrino detection because it is a well-understood process with a known cross-section, which aids in calibration and background rejection. The 2.2 MeV gamma photon is a clear signal that can be correlated in time and position with the preceding interaction, making it a valuable tool for event identification.

The experiment detected the positron (\(e^+\)) produced in this interaction by the light (Cherenkov and scintillation) emitted in the detector. The associated neutron was detected indirectly when it was captured by a proton, producing a 2.2 MeV photon as part of the signature.

A small fraction of the pions and muons decayed in flight before coming to rest. In DIF, the neutrino energy spectrum was broader, but the flux was lower. LSND searched for electron neutrinos (\(\nu_e\)) from pion decay in flight through interactions like:
\begin{align*}
&\nu_e + ^{12}C \rightarrow e^- + X\\
&\overline{\nu}_e + ^{12}C \rightarrow e^+ + X
\end{align*}
The electron energy spectrum from DIF events is expected to be broader than from DAR events, making it more challenging to distinguish neutrino oscillations in DIF. However, the background from conventional neutrino events was expected to be lower.

Most positive pions decayed at rest, producing a well-known flux of muon neutrinos. The challenge was to identify electron neutrinos (\(\nu_e\)) that might be the result of oscillation, while minimizing background events from conventional sources. Negative pions (\(\pi^-\)) and muons that decayed in flight contributed very little to the electron neutrino (\(\nu_e\)) background. Most of the negative pions were captured before they could decay, and any remaining muons were absorbed in the beam stop.

The LSND detector was designed to efficiently recognize electron-like signals associated with neutrino interactions. A major challenge was distinguishing between true oscillation events and background events, particularly those arising from conventional neutrino interactions. The requirement of an electron energy above 36 MeV helped suppress background from charged current interactions involving conventional neutrinos. Additionally, the detection of the 2.2 MeV photon from neutron capture served as a key signature to confirm electron neutrino interactions.
\section{Experimental setup}
\subsection{Scintillation}
The detector was positioned approximately 30 meters from the neutrino source and was shielded by the equivalent of 9 meters of steel. It was surrounded by a liquid scintillator veto shield on all sides except the bottom. The detector itself consisted of a tank containing 167 metric tons of mineral oil $CH_2$ with a small addition of butyl PBD (butyl-phenyl-bipheny-oxydiazole) scintillant at a concentration of 0.031 $\frac{g}{L}$ \cite{aberle_2012}. This dilute mixture enabled the detection of both Cherenkov light and isotropic scintillation light.
\begin{figure}
\centering
\caption{butyl PBD}
\includegraphics[width = 0.4\textwidth, height = 0.13\textwidth]{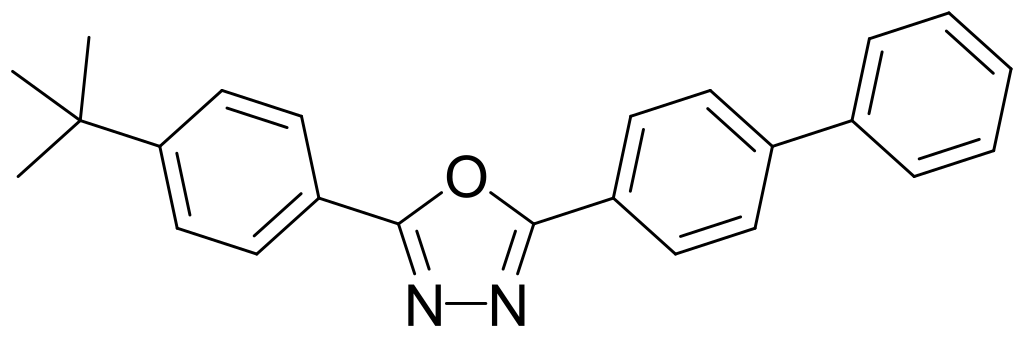}
\end{figure}
\begin{figure}
\centering
\caption{The energy levels of the singlet (S) and triplet (T) states, with their corresponding vibrational levels, are illustrated for a fluorescent molecule. Solid arrows indicate transitions that result in fluorescence or phosphorescence emission, dotted arrows represent non-radiative transitions.}
\includegraphics[width = 0.3\textwidth, height = 0.2\textwidth]{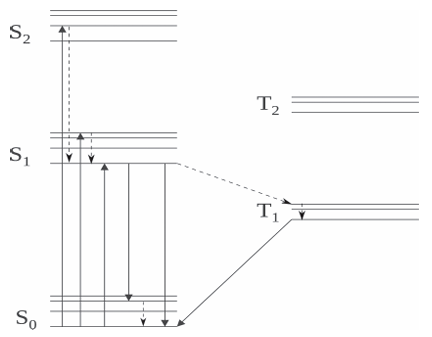}
\end{figure}
In a scintillation cocktail, the solvent molecules absorb part of the energy from an alpha or beta particle. This energy is then transferred between solvent molecules until it reaches a phosphor (a chemical compound that emits light), which absorbs the energy and re-emits it as light.

The number of photons produced is directly proportional to the path length of the beta ($\beta$) particle, which depends on its emission energy. As the $\beta$ particle interacts with solvent molecules, it transfers its energy until it is completely exhausted. The intensity of each light pulse reflects the emission energy, while the number of pulses per second corresponds to the rate of radioactive emissions.

When a single $\beta$ particle passes through the scintillation fluid, it generates multiple, nearly simultaneous light emissions. These photons are detected by the photomultiplier tube (PMT) as a single pulse of energy. The magnitude of this pulse is proportional to the number of photons produced.

The light was detected by 1,220 eight-inch photomultiplier tubes (PMTs), covering approximately 25\% of the inner surface of the tank wall. Each channel recorded data on pulse height and timing. The electronics and data acquisition systems were specifically designed to detect and correlate events occurring at different times.
\subsection{Linear accelerator}
At Los Alamos, the linear accelerator accelerates protons to a specific and consistent energy, which is critical for producing the desired secondary particles (pions and kaons) when the proton beam interacts with a target. These secondary particles decay into neutrinos. The linear accelerator used conventional ion sources to provide protons and H\(^-\) ions, selecting particles on a pulse-by-pulse basis. Protons were accelerated through a transition section into a drift tube linear accelerator (Alvarez type) operating at 201.25 MHz, reaching 50 MeV. They were then injected into a side-coupled linear accelerator structure at 805 MHz, accelerating to 800 MeV. This process repeated at 120 Hz, with all cavities operational to maintain a consistent proton beam energy within 1\% accuracy.

The accelerator often alternated between proton and H\(^-\) beams in interleaved pulses. During much of the experiment, the 120 Hz proton beam was delivered to the experimental area. Each 120 Hz pulse lasted about 600 $\mu$s and featured a substructure of 0.25 ns pulses at 201.25 MHz. The beam current at the accelerator output was typically 1 mA, and the RF frequency remained stable to parts in \(10^7\), ensuring consistent acceleration. Beam phase relative to the RF waveform was optimized for performance, with repetition rates synchronized to the line frequency, varying only 0.1\% daily.

A beam permit signal (H\(^+\)) was issued before acceleration and lasted through delivery, used to verify the timing of neutrino events relative to the accelerator cycle. This signal was essential for beam-on/off subtraction. Beam amplitude varied by less than 1\% over short intervals, and overall availability exceeded 90\%.

\begin{figure}
\centering
\caption{target box plan view}
\includegraphics[width = 0.5\textwidth, height = 0.2\textwidth]{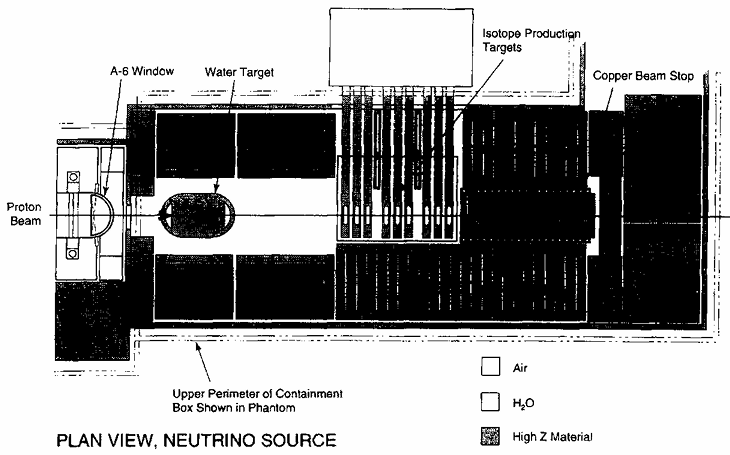}
\end{figure}
\begin{figure}
\centering
\caption{target box elevation view}
\includegraphics[width = 0.5\textwidth, height = 0.2\textwidth]{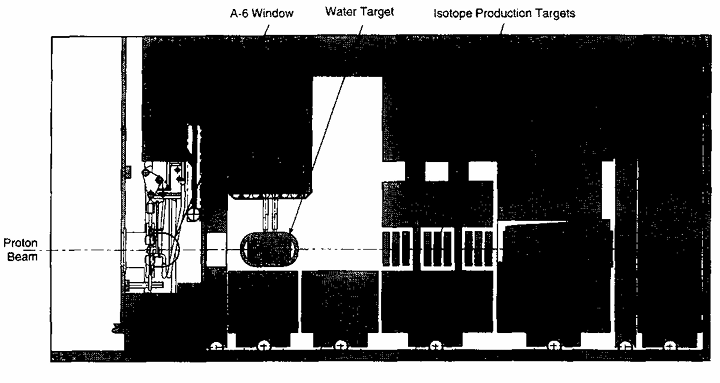}
\end{figure}

Secondary beams were generated at upstream targets A1 and A2 (carbon 3 cm and 4 cm thick) to produce pion and muon beams for other experiments. Pion production and subsequent decay in nearby evacuated enclosures were also potential sources of neutrinos \cite{jansky_2021}. The neutrino flux from A1 and A2 was estimated to be about 1.5\% of that from the primary neutrino target A6. This was based on a production rate 25\% that of A6 and a typical distance four times farther from the detector.

The proton beam started at the left side, went through a water target, and ended in the beam stop. The target is made of an inconel (nickel-chromium-based superalloy) vessel filled with water and fitted with baffles (a device used to restrain the flow of a fluid) to direct flow. Rapid water flow was monitored to prevent boiling during beam passage. The inconel-to-water mass ratio was about 25\% for the proton beam, slightly higher for secondary pions produced at angles. Water-cooled iron shielded the entire region, which was hermetically sealed to prevent neutron background and air activation hazards.

Pion decay occurred in the shielded region downstream of the target. Negative pion decay was suppressed, as stopping \(\pi^-\) were absorbed, and \(\mu^-\) from in-flight \(\pi^-\) decay stopped in iron, copper, or water-cooling channels, with about 9\% stopping in the water target. Typical pions had a momentum of 200 MeV/c and a \(\beta\gamma c\tau\) of 12 m ($\beta=\frac{v}{c}, \gamma$ is lorentz factor $\gamma=1/(\sqrt{1-\frac{v^2}{c^2}}), \tau$ is particle's proper lifetime in rest frame), leading to about 3.4\% decaying in flight. Upstream isotope production targets (A1 and A2) and their shielding were accounted for in neutrino flux simulations. These targets were mounted in enclosures with hermetic shielding, except for vacuum pipes allowing four secondary beams to exit from each side.

\section{Theoretical modeling and simulation}
To study the reactions $\nu_e + p \rightarrow e^- + n,\; \overline{\nu}_e + p \rightarrow e^+ + n$, and $n + p \rightarrow d + \gamma$, the FLUKA Monte Carlo code is used to simulate particle transport and interactions with scintillator materials. It supports neutron transport in the thermal range up to 20 TeV, hadronic charged particles from 1 keV to 20 TeV, and heavy charged particles up to \(10^4\) TeV \cite{ghavami_2024}. FLUKA employs various physical models and algorithms for accurate particle transport. FLUKA uses two internal thresholds for neutron transport: one for high-energy and one for low-energy neutrons. High-energy neutrons transition to multigroup transport at an energy boundary defined by the cross-section library. For the 260-group library included with the code, this boundary is 20 MeV. Continuous (pointwise) cross-sections are integrated for isotopes like \(^{1}\text{H}\), \(^{2}\text{H}\), \(^{3}\text{He}\), \(^{4}\text{He}\), and \(\text{C}\), automatically adjusted for specified temperature conditions.

\subsection{Response function}
For the reactions \( \nu_e + p \rightarrow e^- + n \) and \( \overline{\nu}_e + p \rightarrow e^+ + n \), mono-energetic neutrons are generated to simulate the secondary neutrons produced. The energy and trajectory of these neutrons are specified using the BEAM and BEAMPOS cards. Secondary particles, including neutrons and \( e^\pm \), interact with the scintillator material. The AUXSCORE card identifies whether the secondary particle is an electron, positron, or neutron, and the EVENTBIN card calculates the energy deposited in the scintillator by \( e^- \) or \( e^+ \), which produces scintillation light. The TCQUENCH card estimates the scintillation light output from \( e^\pm \), which is proportional to their energy and detected as the primary signal. The neutron produced in the interaction will scatter within the detector and eventually be captured by a hydrogen nucleus in the scintillator, resulting in the reaction \( n + p \rightarrow d + \gamma \).

\begin{figure}
\centering
\caption{simulated response function}
\includegraphics[width = 0.35\textwidth, height = 0.23\textwidth]{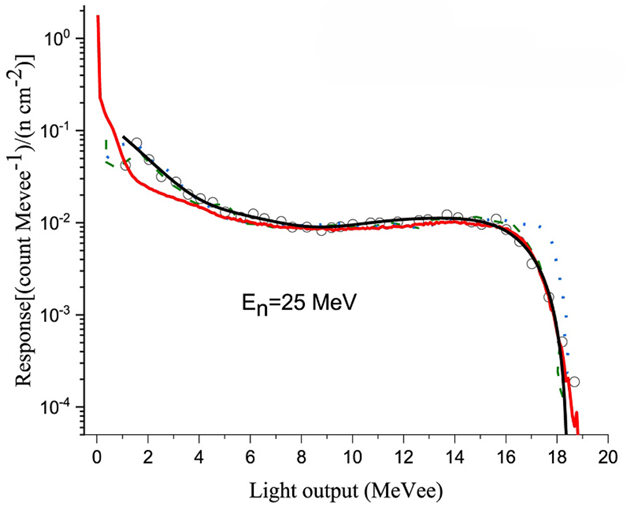}
\end{figure}
\begin{figure}
\centering
\caption{phenyl xylylethane (PXE) attentuation length vs wavelength, with 2,5-diphenyloxazole (PPO) and 4-bis-(2-methyl
styryl)benzene (bis-MSB)}
\includegraphics[width = 0.35\textwidth, height = 0.23\textwidth]{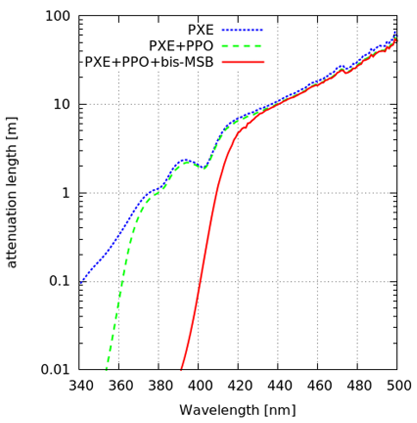}
\end{figure}

The response function have units (count MeVee)\(^{-1}\)/(\(\text{n cm}^{-2})\). The y-axis response function refers to the number of detected events or counts in the scintillator as a function of light output. The detector records these counts based on the light signals generated by neutron interactions in the scintillator material. The term "MeVee" stands for "MeV electron-equivalent," which is a measure of light output in the scintillator. It represents the amount of light produced as if the energy were deposited by an electron with a corresponding energy measured in MeV. MeVee indicates that the response is normalized per unit of light output (electron-equivalent energy). The term (\(\text{n cm}^{-2}\)) represents the neutron fluence, which is the number of neutrons passing through a unit area, in neutrons per square centimeter. The denominator indicates that the response is normalized per unit neutron fluence, making the y-axis quantity a measure of the detector’s sensitivity to neutrons. The y-axis shows the normalized response of the scintillator as a function of light output. The units indicate that this response is given in terms of counts per unit of light output (MeVee) per neutron fluence (\(\text{n cm}^{-2}\)), allowing comparison of the detector’s performance across different neutron energies and light output values while accounting for variations in neutron fluence.

The response function decreases as the light output increases due to the physics of neutron interactions and the nature of the scintillator's light production mechanisms. At lower light output, corresponding to lower-energy charged particles produced by neutron interactions (such as protons from elastic scattering), the response is higher because these interactions are more probable. Elastic scattering dominates at lower neutron energies, and the scintillator produces a strong response due to the relatively high ionization energy loss (\(dE/dx\)) of the recoiling protons.

As the light output increases, corresponding to higher-energy charged particles, the response decreases because higher-energy interactions often involve less frequent processes, such as inelastic neutron-carbon interactions or multiple scattering events, which produce less light per unit of deposited energy. Additionally, scintillator quenching effects (described by Birks' law) reduce the light yield for high-energy particles. The scintillator becomes less efficient at converting the deposited energy into light as the energy density along the particle’s track increases, leading to a diminishing response at higher light outputs.

In the neutrino interactions, the primary signal corresponds to the scintillation light from \( e^- \) or \( e^+ \), while the secondary signal corresponds to the delayed neutron-capture signal (\( n + p \rightarrow d + \gamma \)) within the scintillator. The timing correlation between these signals allows discrimination of neutrino events from background signals. 

The simulation models the energy deposition and light yield for each of these reactions to understand the detector response to \( \nu_e \), \( \overline{\nu}_e \), and neutron capture events, distinguish between different interactions based on light output and timing, and verify that the scintillator's response matches experimental observations, enabling more precise interpretation of neutrino experiments like LSND.

\subsection{$\frac{dL}{dE}$ light output per energy}
The light yield \(L\) in a scintillator is given by the relationship:
\begin{align*}
&\frac{dL}{dx}=S\frac{dE}{dx}\left(1+KB\frac{dE}{dx}\right)^{-1}\\
&\frac{dL}{dE}=S\left(1+KB\left(\frac{dE}{dx}\right)+C\left(\frac{dE}{dx}\right)^2\right)^{-1}
\end{align*}
where $E$ is the energy, $x$ is the particle path length, \(S\) is the scintillation efficiency, \(\frac{dE}{dx}\) is the specific energy loss of the particle per path length, \(K\) is the probability of quenching, \(B\) is a constant of proportionality linking the local density of ionized molecules along the particle's path to the specific energy loss. The product \(KB\), known as Birks' coefficient \cite{birks_1951}, acts as a single parameter with units of distance per energy. Its value depends on the scintillating material. For polystyrene-based scintillators, $kB$ = 0.126mm/MeV, while for polyvinyltoluene-based scintillators, \(kB\) ranges from \(1.26\) to \(2.07 \times 10^{-2} \, \text{g MeV}^{-1} \text{cm}^{-2}\).

Birks suggested that the nonlinearity in light yield arises from recombination (ionized molecules (ions and electrons) created along the particle's track recombine without producing scintillation light) and quenching (loss of energy through non-radiative processes rather than light emission) effects between excited molecules and the surrounding substrate. Birks' law has been primarily tested with organic scintillators.

To simulate the neutron response function of a scintillation detector in the energy range of 25 MeV–800 MeV, the scintillating material selected is BC501A. This material has a cylindrical geometry with a diameter and height of 12.7 cm and is characterized by the molecular formula \(\text{C}_6\text{H}_4(\text{CH}_3)_2\) and a density of \(0.874 \, \text{g/cm}^3\). The detector is exposed to mono-energetic neutrons, whose energy, position, and trajectory are specified using the BEAM and BEAMPOS cards. The TCQUENCH card computes the light output by secondary charged particles using the scintillator’s light yield equation, \(\frac{dL}{dE}\), where \(dL\) is the differential light output and \(dE\) is the differential energy loss.

\begin{align*}
&\frac{dE}{dx}=\exp(b_0+b_1\log E+b_2(\log E)^2+b_3(\log E)^3)\\
&L=a_1E-a_2(1-\exp(-a_3E))\\
&\frac{dL}{dE}=S(1+KB(\exp(b_0+b_1\log E+b_2(\log E)^2+b_3(\log E)^3)))\\
&\quad+C(\exp(b_0+b_1\log E+b_2(\log E)^2+b_3(\log E)^3)^2)^{-1}
\end{align*}

\begin{figure}
\centering
\caption{proton $\frac{dL}{dE}$ light output per energy vs energy graph}
\includegraphics[width = 0.35\textwidth, height = 0.23\textwidth]{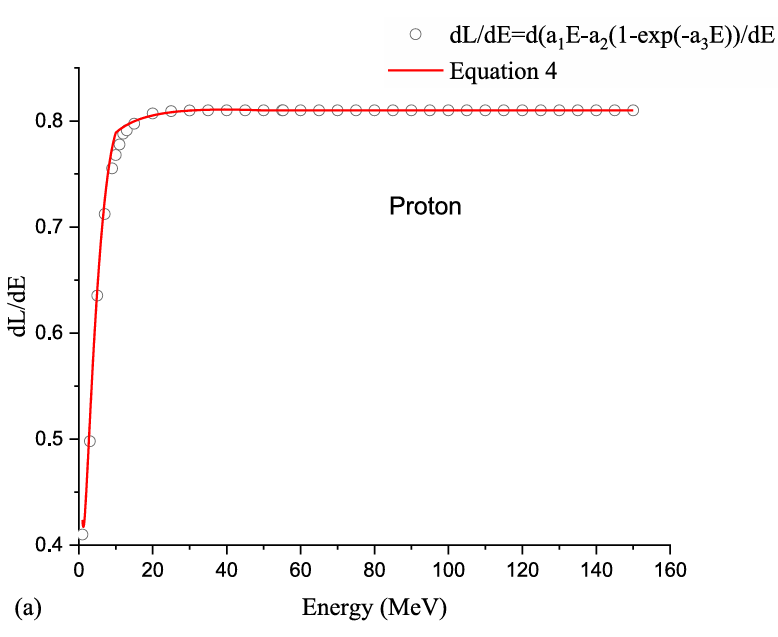}
\end{figure}
\begin{figure}
\centering
\caption{deuteron $\frac{dL}{dE}$ light output per energy vs energy graph}
\includegraphics[width = 0.35\textwidth, height = 0.23\textwidth]{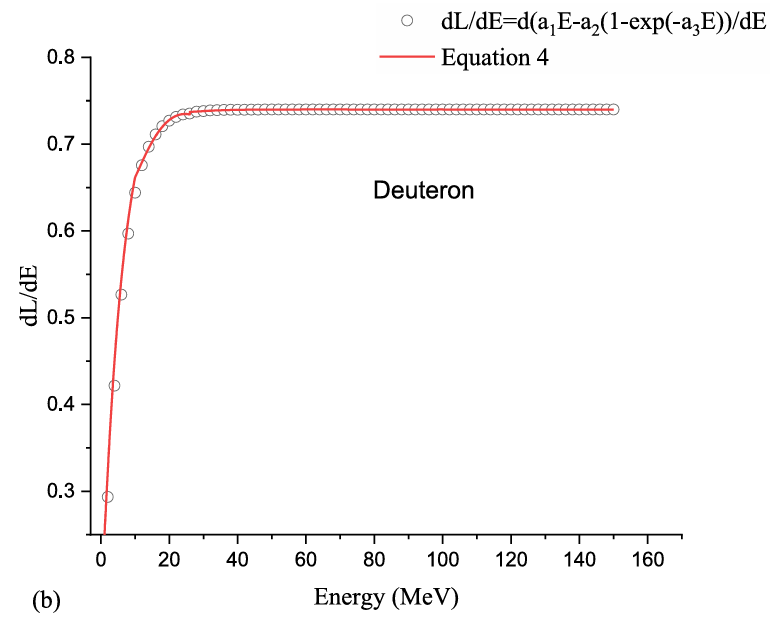}
\end{figure}
\begin{figure}
\centering
\caption{alpha $\frac{dL}{dE}$ light output per energy vs energy graph}
\includegraphics[width = 0.35\textwidth, height = 0.23\textwidth]{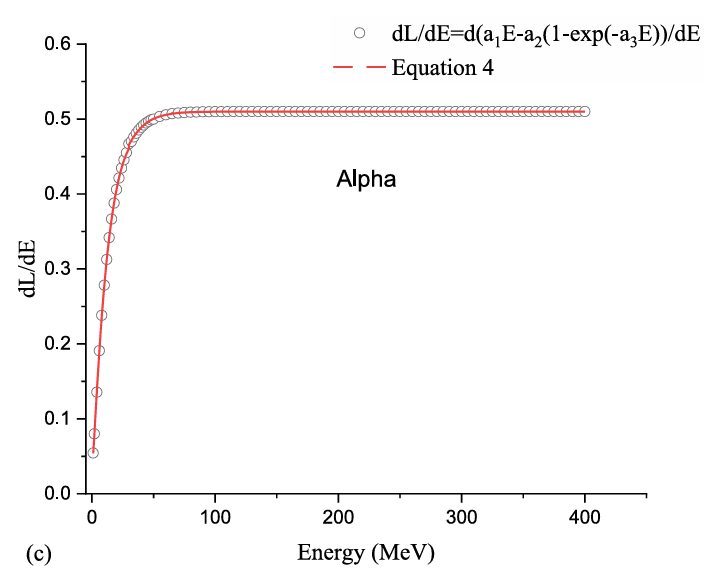}
\end{figure}
\begin{figure}
\centering
\caption{Tritium $\frac{dL}{dE}$ light output per energy vs energy graph}
\includegraphics[width = 0.35\textwidth, height = 0.23\textwidth]{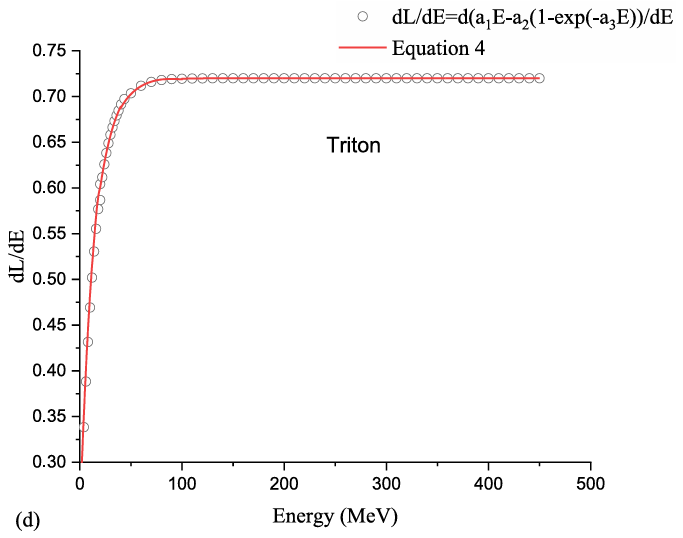}
\end{figure}

The \( \frac{dL}{dE} \) curve illustrates how efficiently the scintillator converts the deposited energy of secondary charged particles (such as electrons, positrons, or recoil nuclei) into light. This efficiency varies with particle energy, and the graph helps to identify regions of linearity or non-linearity in light yield. The curve reflects the effects of quenching as described by Birks' law. At higher energy deposition rates (or stopping powers), the light output deviates from linearity due to recombination and quenching processes. The graph helps quantify this behavior and evaluate parameters like Birks' coefficient (\( KB \)). Different particles produce distinct \( \frac{dL}{dE} \) curves. Analyzing these differences allows for the identification and discrimination of particle types based on their energy deposition and resulting light yield. The \( \frac{dL}{dE} \) plot provides a benchmark for validating simulation models like FLUKA. For reactions \( \nu_e + p \rightarrow e^- + n \), \( \overline{\nu}_e + p \rightarrow e^+ + n \), and \( n + p \rightarrow d + \gamma \), the graph helps relate the energy of the secondary charged particles (electrons, positrons, or recoils) to the light output, ensuring the detector response is consistent across the expected energy range.

\subsection{Neutrino flux calculation}
\begin{figure}
\centering
\caption{neutrino flux from pion and muon decay at rest}
\includegraphics[width = 0.35\textwidth, height = 0.23\textwidth]{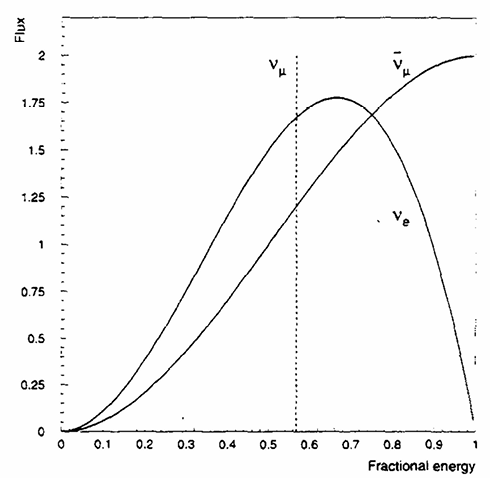}
\end{figure}
\begin{figure}
\centering
\caption{$\nu_e$ flux vs energy for $\pi^+$ decay}
\includegraphics[width = 0.33\textwidth, height = 0.23\textwidth]{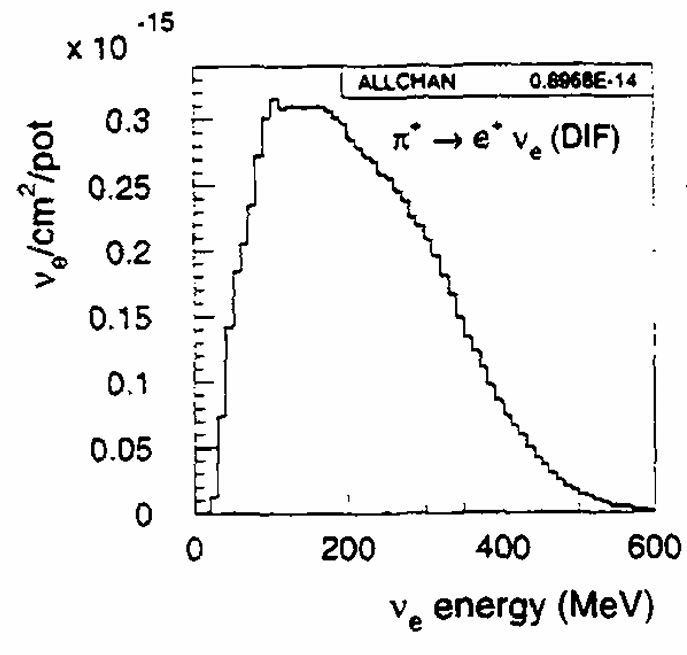}
\end{figure}
\begin{figure}
\centering
\caption{$\nu_e$ flux vs energy for $\mu^+$ decay}
\includegraphics[width = 0.33\textwidth, height = 0.23\textwidth]{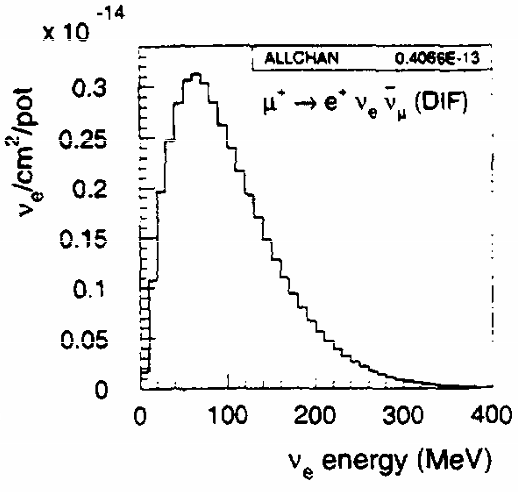}
\end{figure}
\begin{figure}
\centering
\caption{$\overline{\nu}_e$ flux vs energy for $\pi^-$ decay}
\includegraphics[width = 0.33\textwidth, height = 0.23\textwidth]{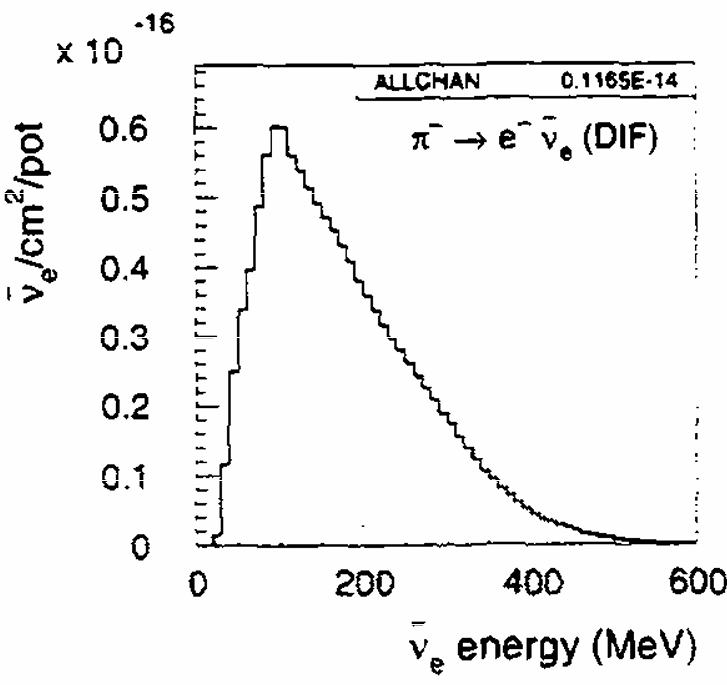}
\end{figure}
\begin{figure}
\centering
\caption{$\overline{\nu}_e$ flux vs energy for $\mu^-$ decay}
\includegraphics[width = 0.33\textwidth, height = 0.23\textwidth]{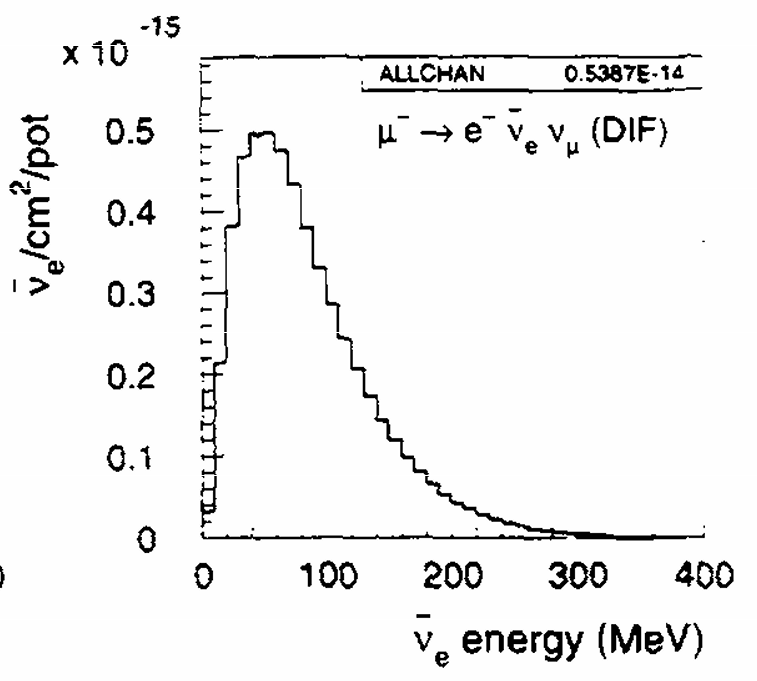}
\end{figure}
The graphs for the neutrino flux from pion and muon decays increase to a peak value and then decrease to zero because of the energy distribution of neutrinos resulting from the kinematics and phase space of the decay processes. The $\nu_\mu$ curve increases within the interval $x = 0$ to $x = 0.57$. For pion decay in flight (DIF), such as $\pi^+ \to \mu^+ + \nu_\mu$, the resulting neutrinos ($\nu_\mu$) have a wide range of energies because of the momentum of the decaying pion \cite{burman_1990}. Subsequently, muons from pion decays can decay in flight ($\mu^+ \to e^+ + \nu_e + \overline{\nu}_\mu$), producing $\nu_e$ and $\overline{\nu}_\mu$.

The neutrino flux increases initially because more pions decay into neutrinos with moderate energies due to the phase space available for such decays. At low energies, fewer neutrinos are produced due to phase space constraints. The flux decreases after reaching a peak because fewer pions and muons can produce neutrinos with higher energies as their decay kinematics limit the maximum neutrino energy. This results in the suppression of flux at high energies.

For muon decay at rest (DAR), such as $\mu^+ \to e^+ + \nu_e + \overline{\nu}_\mu$, the neutrinos are distributed over a specific energy range due to the three-body decay nature. At low neutrino energies, the flux is lower because the phase space for producing neutrinos with very low energy is small. The flux increases and peaks at intermediate energies due to the balance of available phase space and kinematic constraints. At higher energies, the flux decreases to zero because the maximum neutrino energy is limited by the decay kinematics, particularly by the fixed mass difference between the muon and its decay products.

\subsection{Pion production cross sections}
Cross section value is a measure of the probability of a pion being produced in a specific interaction. A larger cross section implies a higher probability of the interaction happening, meaning more pions will be produced per incident particle under the same conditions.\\
Production rate = (Cross section) $\times$ (Incident particle flux) $\times$ (Target density)
\begin{align*}
R_\pi=\Phi_p\sigma_\pi N_t
\end{align*}
where \(R_\pi\) is the pion production rate (number of pions produced per second or event), \(\Phi_p\) is the proton flux (number of protons incident on the target per unit area per unit time, typically given in \(\text{protons}/\text{cm}^2/\text{s}\)), \(\sigma_{\pi}\) is the pion production cross-section (typically in units of barns, where \(1 \text{ barn} = 10^{-24} \, \text{cm}^2\)), and \(N_t\) is the number of target nuclei per unit area, which depends on the material density and thickness.
\begin{align*}
&\frac{d^2\sigma}{d\Omega_\pi dT_\pi}=Amp(\theta)e^{-\left(\frac{\overline{T}-T_\pi}{\sqrt{2}\sigma(\theta)}\right)^2}\\
&\quad\times(1+e^{\frac{T_\pi-T_F}{B}})^{-1}(\mu b MeV^{-1}sr^{-1})\\
&\sigma_\pi=\iint\frac{d^2\sigma}{d\Omega_\pi dT_\pi}dT_\pi d\Omega_\pi
\end{align*}

The geometrical configuration of the target is managed using the external geometry package of the General Monte Carlo Code for Neutron and Photon Transport (MCNP). This is specified through an input file.

The parameters Amp($\theta$), \( T \), and \( a \) were based on experimental data. These were derived at two proton energies, \(T_p = 585\, \text{MeV} \) and \( T_p = 730 \, \text{MeV} \), for hydrogen, carbon, copper, and lead. The resulting values were expressed in simple forms to generalize across the range of variables. The functional dependence on \( T_p \) is straightforward, appearing in the exponent of the Gaussian and the high-energy cut-off factor. Analytic forms are sufficient to describe the angular dependence of the Gaussian energy parameters \(\overline{T}\) and \( \sigma \):

\begin{align*}
&\overline{T}(\theta)=48+330e^{-\frac{\theta}{T_A}}\\
&\sigma(\theta) = \sigma_A e^{-\frac{\theta}{85}}\\
&T_A(Z, T_p) = \frac{T_A(Z, 730)(T_p - 585) - T_A(Z, 585)(T_p - 730)}{730 - 585}\\
&\sigma_A(Z, T_p) = \frac{\sigma_A(Z, 730)(T_p - 585) - \sigma_A(Z, 585)(T_p - 730)}{730 - 585}
\end{align*}
The amplitude \( \text{Amp}(\theta)\) depends on the material, proton energy, and scattering angle. To account for this nontrivial dependence, it has been parameterized using 5-splines over the allowed angular region \( 0^\circ < \theta < 180^\circ \).
\begin{align*}
&Amp(\theta)=Norm(Z)\times\sum_{n=1}^5a_nB_n\\
&a_1=\min\left(27-4\left(\frac{730-T_p}{730-585}\right)^2, 27\right)\\
&a_2=18.2, a_3=8, a_4=13+(Z-12)/10\\
&a_5=9 + (Z - 12)/10 - (T_p - 685)/20\\
&Norm(Z)=\sum_{m=0}^3c_m(\ln Z)^mZ^{1/3}\\
&c_0=0.8851, c_1=-0.1015, c2 = 0.1459, c3 = -0.0265\\
\end{align*}
\begin{figure}
\centering
\caption{$\pi^+$ cross sections}
\includegraphics[width = 0.33\textwidth, height = 0.35\textwidth]{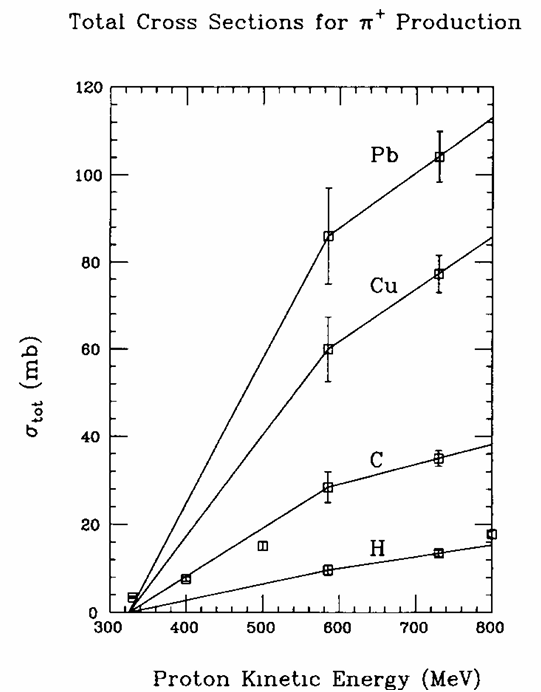}
\end{figure}
\begin{figure}
\centering
\caption{$\pi^+$ production from hydrogen, x-axis is $T_{\pi}(MeV)$ (from 0 to 600), y-axis is $\sigma(\mu b/MeV/sr)$ (from 0 to 30)}
\includegraphics[width = 0.4\textwidth, height = 0.25\textwidth]{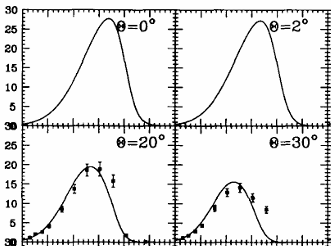}
\includegraphics[width = 0.4\textwidth, height = 0.25\textwidth]{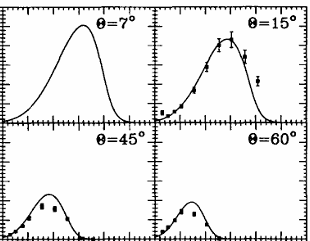}
\end{figure}
The total cross section, obtained by integrating the parameterized differential cross section, reproduces published cross sections to within approximately 10\%. The absolute cross section is treated independently to enable flexibility in assessing overall normalization effects. It is assumed that the total cross section varies linearly with \( T_p \). For \( T_p > 585 \, \text{MeV} \), the curve is determined by the total cross sections at \( T_p = 730 \, \text{MeV} \) and \( T_p = 585 \, \text{MeV} \). Below the break point of \( T_p = 585 \, \text{MeV} \), the total cross section decreases linearly to zero at \( T_p = 325 \, \text{MeV} \).

The total cross section dependence on atomic number \( Z \) for pion production has been modeled. For \( Z > 11 \), a \( Z^{1/3} \) dependence is applied, while for \( Z < 12 \), an additional linear term in \( Z \) is included to improve agreement with data for lighter elements, such as helium. The total cross section for hydrogen follows a specialized case. The expressions for the cross section are
\[
\sigma(Z > 11) = c_0(T_p) Z^{1/3}
\]
\[
\sigma(Z < 12) = c_2(T_p) \left(\frac{Z}{6}\right)^{1/3} \left(0.77 + 0.039 Z\right)
\]
\[
\sigma(Z = 1) = \sigma_1(T_p)
\]

The data at \( T_p = 730 \, \text{MeV} \) are presented, and the parameterization reproduces these data. However, deviations occur: underestimation at \( T_p = 585 \, \text{MeV} \) and scattering angles around \( 50^\circ \), and overestimation at \( T_p = 730 \, \text{MeV} \) and \( 20^\circ \). These discrepancies may reflect either inherent limitations in the parameterization or inconsistencies in the data. Validation of predicted cross sections at higher energies and forward angles shows reasonable agreement with independent measurements not used to set parameters. Overall, this parameterization effectively predicts pion production cross sections across all materials, proton energies below \( 800 \, \text{MeV} \), and production angles.

\begin{figure}
\centering
\caption{$\pi^+$ production from carbon, x-axis is $T_{\pi}(MeV)$ (from 0 to 600), y-axis is $\sigma(\mu b/MeV/sr)$ (from 0 to 40)}
\includegraphics[width = 0.4\textwidth, height = 0.25\textwidth]{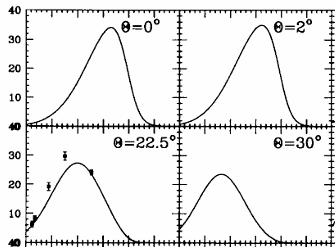}
\includegraphics[width = 0.4\textwidth, height = 0.25\textwidth]{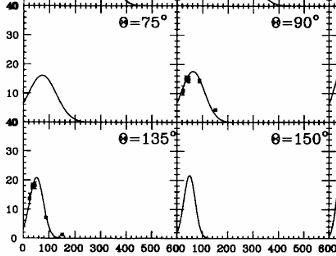}
\end{figure}

\subsection{Scintillation}
Liquid scintillation processes rely on concepts of electronic excitation, energy transfer, and radiative transitions within aromatic molecules and associated fluorophores. The excitation of aromatic solvent molecules by ionizing particles is governed by the time-independent Schrodinger equation, which describes the energy levels of the system. The radiative decay rate of excited states, such as fluorescence, is determined by Einstein coefficients. For spontaneous emission, the rate \(k_f\) is proportional to the transition probability: $k_f = A_{ij}$, where \(A_{ij}\) denotes the spontaneous emission coefficient between electronic states \(i\) and \(j\). Energy transfer from the solvent to a fluor or wavelength shifter often occurs via Förster Resonance Energy Transfer (FRET), which depends on the spectral overlap and distance between the donor and acceptor molecules:

\begin{align*}
k_{\text{FRET}} \propto \frac{1}{R^6} \frac{\int f_D(\lambda) \cdot f_A(\lambda) \, d\lambda}{n^4}
\end{align*}

where \(R\) is the donor-acceptor separation distance, \(f_D(\lambda)\) and \(f_A(\lambda)\) are the fluorescence and absorption spectra, respectively, and \(n\) is the refractive index of the medium.

The scintillation efficiency is also characterized by the quantum yield (QY), which measures the fraction of absorbed energy converted into emitted light:

\begin{align*}
QY = \frac{\text{Number of photons emitted}}{\text{Number of photons absorbed}}
\end{align*}

Self-absorption of scintillation light is influenced by the spectral overlap integral between absorption and emission spectra:

\begin{align*}
J = \int S_E(\lambda) S_A(\lambda) \, d\lambda
\end{align*}

where \(S_E(\lambda)\) and \(S_A(\lambda)\) represent the emission and absorption spectra, respectively. Minimizing this overlap is essential for enhancing the transparency of LSs.

In addition to scintillation, LSs can produce Cerenkov radiation, which arises when a charged particle travels faster than the speed of light in the medium: $v > \frac{c}{n}$. The angle of Cerenkov emission is given by: $\cos \theta_C = \frac{1}{\beta n}$, where \(\beta = v/c\) and \(n\) is the refractive index. Let \(\alpha\) be the fine-structure constant, the number of photons emitted per unit wavelength and distance due to Cerenkov radiation is:

\begin{align*}
\frac{d^2N}{d\lambda dx} = \frac{2\pi \alpha}{\lambda^2} \left(1 - \frac{1}{\beta^2 n^2}\right)
\end{align*}

Light propagation in LSs is characterized by the attenuation length, which describes the exponential decrease in intensity due to absorption and scattering:

\begin{align*}
I(x) = I_0 e^{-x / L_{\text{att}}}
\end{align*}

where \(L_{\text{att}}\) is the attenuation length. Additionally, non-radiative processes such as internal conversion (IC) and intersystem crossing (ISC) play a role in de-excitation. These processes depend on Franck–Condon factors and thermal activation:

\begin{align*}
k_{\text{IC}} \propto \sum_v |\langle \psi_v^f | \psi_v^i \rangle|^2 e^{-\Delta E / k_BT}
\end{align*}

To examine the different materials in the liquid scintillator, molecular orbital (MO) theory can be used to understand the bonding and electronic structure of organometallic ligands, particularly in the context of metal-ligand interactions. At the core of MO theory is the concept of atomic orbitals (AOs) combining to form molecular orbitals (MOs). The molecular orbitals can be classified into bonding, anti-bonding, or non-bonding orbitals, depending on the symmetry and overlap of the AOs involved. In organometallic complexes, the ligand donor orbitals (such as lone pairs on oxygen or nitrogen atoms) interact with the metal's vacant or partially filled orbitals to form bonding and anti-bonding MOs.

The combination of atomic orbitals from two atoms \(A\) and \(B\) leads to the formation of molecular orbitals, which can be expressed as a linear combination of atomic orbitals (LCAO). The general form of the molecular orbital is given by: $\psi_{\text{MOs}} = c_A \psi_A + c_B \psi_B$, where \(c_A\) and \(c_B\) are coefficients that describe the contribution of the atomic orbitals \(\psi_A\) and \(\psi_B\) to the molecular orbital. These coefficients are determined by the overlap integral between the atomic orbitals and the energy levels of the atoms involved in the bonding.

The energy of a molecular orbital is determined by the energies of the atomic orbitals and the overlap between them. For two atomic orbitals \(A\) and \(B\), the molecular orbital energy can be approximated by the following equation: $E_{\text{MO}} = \frac{E_A + E_B}{2} \pm \frac{1}{2} \sqrt{(E_A - E_B)^2 + 4V^2}$, where \(E_A\) and \(E_B\) are the energies of the atomic orbitals and \(V\) is the orbital overlap integral. The positive sign corresponds to the anti-bonding molecular orbital, while the negative sign corresponds to the bonding molecular orbital.

The bond order of a molecule or complex is an important measure of its stability and is given by the difference between the number of electrons in bonding and anti-bonding orbitals. The bond order can be expressed as: $\text{Bond Order} = \frac{1}{2} (N_{\text{bonding}} - N_{\text{anti-bonding}})$, where \(N_{\text{bonding}}\) and \(N_{\text{anti-bonding}}\) are the numbers of electrons in the bonding and anti-bonding molecular orbitals, respectively. A higher bond order indicates a more stable complex, which is crucial for understanding the stability of metal-ligand complexes in applications like metal-loaded scintillators.

In metal-ligand complexes, the ligand field created by the surrounding ligands splits the degenerate d-orbitals of the metal ion into different energy levels. This splitting is described by the ligand field theory, and the energy difference between the split orbitals is given by the ligand field splitting energy, \(\Delta E\). For the \(d\)-orbitals, the splitting energy can be expressed as: $\Delta E = E_{\text{eg}} - E_{\text{t2g}}$, where \(E_{\text{eg}}\) and \(E_{\text{t2g}}\) are the energies of the \(e_g\) and \(t_{2g}\) orbitals, respectively. This splitting affects the electronic configuration of the metal ion and is critical for understanding the optical properties and stability of metal-loaded liquid scintillators.

Charge transfer processes, including ligand-to-metal charge transfer (LMCT) and metal-to-ligand charge transfer (MLCT), also play an important role in organometallic chemistry. The energy associated with these charge transfer transitions can be calculated as: $E_{\text{CT}} = E_{\text{MLCT}} - E_{\text{LMCT}}$, where \(E_{\text{MLCT}}\) and \(E_{\text{LMCT}}\) are the energies for metal-to-ligand and ligand-to-metal charge transfer, respectively. These transitions influence the electronic properties of the material, particularly in metal-loaded scintillators used for particle detection in experiments like LSND.

The formation of molecular orbitals and the associated electronic structure are influenced by the back donation mechanism, especially in systems involving transition metals and ligands with $\pi$-acceptor properties \cite{brown_2004}. The energy levels of the molecular orbitals formed by this interaction affect the material’s ability to absorb and emit light, which is essential for the performance of scintillators in experiments detecting low-energy particles such as neutrinos.

Liquid scintillators (LSs) are commonly employed as target materials for detecting reactor neutrinos. Researchers have investigated the incorporation of metals such as gadolinium (Gd) and lithium (Li) into these neutrino targets to enhance their capability to distinguish signals from background noise \cite{maddalena_2019}. Studies have also focused on ensuring that liquid scintillators are environmentally friendly, transparent, and stable for long-term use in data collection. To further improve background discrimination, di-isopropyl naphthalene (DIN, C$_{16}$H$_{20}$)-infused LSs have been developed for short-baseline neutrino experiments.

The scintillation process in organic fluorescent materials arises from the energy loss of incident particles. Charged particles primarily lose energy in a medium through interactions with atomic electrons or nuclei. This energy loss can occur via direct energy transfer through continuous Coulomb interactions, leading to a loss of kinetic energy. Alternatively, indirect energy transfer occurs when neutral particles, such as photons or neutrons, gain energy from secondary charged particles via Compton scattering or nuclear reactions. The energy loss of charged particles in a medium is characterized by the Bethe–Bloch formula, which provides the stopping power of a medium for a given particle. Stopping power refers to the rate of energy loss per unit length as the particle travels through the medium. The Bethe–Bloch formula is given by:
\begin{align*}
\frac{dE}{dx} &= \frac{4 \pi e^4 z^2}{m_0 v^2} N Z 
\left[
\ln\left(\frac{2 m_0 v^2}{I}\right) - \ln\left(1 - \frac{v^2}{c^2}\right) - \frac{v^2}{c^2}
\right]
\end{align*}
$\frac{dE}{dx}$ represents the energy loss per unit distance traveled by the particle. $v$ and $z e$ denote the particle's velocity and charge, respectively. $N$ and $Z$ are the number density and atomic number of the medium. $I$ refers to the average excitation energy of the medium. $m_0$ is the rest mass of an electron, and $c$ is the speed of light.

This formula describes the energy loss caused by ionization and excitation of the medium by the charged particle. It is valid for high-energy charged particles, including electrons, protons, and alpha particles. The energy loss arises from Coulomb interactions between the charged particle and the atomic electrons or nuclei within the medium. In a liquid scintillator (LS), the energy loss of particles is detected through the scintillation process. The energy transferred to the medium excites the scintillator molecules, resulting in the emission of photons. These photons are then captured by photomultiplier tubes (PMTs), which convert them into electrical signals. The resulting signals can be analyzed to measure the energy loss of the particles.

As ionizing particles pass through an LS, part of their energy excites the aromatic solvent molecules within the scintillator. This excitation causes electron delocalization in the $\pi$-bonds of the benzene rings or phenyl groups. After this initial excitation, a series of radiationless transitions occur, populating the first excited electronic state of the molecules. These transitions typically take place within a very short time frame, around $10^{-11}$ seconds. Liquid scintillators use aromatic compounds as both solvents and solutes.

In an isolated fluorescent molecule, the transition from the first excited singlet state back to the ground state results in photon emission, a process known as fluorescence. This emission occurs rapidly, typically within a range of a few nanoseconds to tens of nanoseconds. In some cases, transitions to triplet states can also take place, which may result in longer-lasting emission, extending up to milliseconds. This prolonged emission is termed phosphorescence. The selection of solvent is crucial, as it directly affects key properties of the liquid scintillator (LS), including light yield (LY), attenuation length, and decay time.

For a single aromatic solvent molecule, the majority of the fluorescent radiation is emitted from the aromatic ring as scintillation light in an organic liquid. However, the presence of additional molecules, especially wavelength shifters, can modify the light-production mechanism. The solvent in a liquid scintillator (LS) primarily serves as the medium for interaction with ionizing particles. Traditionally, benzene (C$_6$H$_6$) derivatives have been the solvents of choice, with compounds like toluene (C$_7$H$_8$), xylene (C$_8$H$_{10}$), and cumene ($C_9H_{12}$) being commonly used \cite{burman_parameter_1989}. Aromatic compounds are preferred due to their intrinsic fluorescent properties, which provide the necessary energy levels for transitions during the scintillation process. Recently, linear alkyl benzenes (LABs, C$_n$H$_{2n+1}$–C$_6$H$_5$, with $n\in[10,13]$) have gained prominence as a solvent, owing to a combination of factors such as their safety, compatibility with other scintillator components.

The LSND experiment played a crucial role in exploring the possibility of neutrino oscillations by searching for the appearance of electron neutrinos in a predominantly muon neutrino beam. By detecting neutrino interactions in a large tank filled with mineral oil, LSND provided evidence that could challenge the Standard Model, suggesting the possible existence of sterile neutrinos or other new physics. The results, while controversial, paved the way for further experiments like MiniBooNE to explore similar anomalies and refine our understanding of neutrino oscillations.

\section{Conclusion}
The LSND experiment has significantly advanced the understanding of neutrino oscillations and the underlying physics of neutrino interactions. Theoretical modeling and simulation framework guided both experimental design and data interpretation. Using the FLUKA Monte Carlo code, we simulated complex particle interactions and transport mechanisms in the liquid scintillator. These simulations included detailed modeling of neutrino interactions, neutron capture, and the resulting scintillation light output. The incorporation of Birks' law enabled accurate predictions of nonlinear light yield, while detailed calculations of neutrino fluxes from DAR and DIF processes provided a nuanced understanding of the energy distributions of neutrinos.

Simulations of pion production cross-sections and particle-specific light output efficiency ensured a comprehensive evaluation of the detector's performance. The detector response function, which related light output to neutron fluence, highlighted the scintillator's sensitivity to low-energy interactions and its capability to discriminate against background signals. The ability to simulate these processes accurately proved essential in validating the experimental results and reinforcing the evidence for neutrino oscillations. The LSND findings underscored the potential for new physics, including the possibility of sterile neutrinos. The experiment also demonstrated the importance of integrating advanced modeling and simulation techniques in neutrino research, setting a high standard for future experiments. By bridging experimental observations with theoretical predictions, LSND has laid a robust foundation for ongoing exploration into the fundamental properties of neutrinos and their role in the universe.
\bibliographystyle{elsarticle-num}
\bibliography{reference}
\end{document}